\documentclass[conference]{IEEEtran}
\IEEEoverridecommandlockouts
\usepackage{cite}
\usepackage{amsmath,amssymb,amsfonts}

\usepackage{algorithm}
\usepackage[noend]{algpseudocode}

\usepackage{subcaption} 

\usepackage{graphicx}

\usepackage{textcomp}
\usepackage{xcolor}

\usepackage{hyperref}

\usepackage{pgfplots,pgfplotstable}
\usetikzlibrary{pgfplots.groupplots}
\pgfplotsset{compat=1.17}

\usepackage{tikz}
\usepackage{adjustbox}
\usepackage{amsthm} 
\usepackage{color, colortbl}
\usepackage{makecell, cellspace}
\usepackage{array,multirow}

\newcommand{\pie}[1]{%
\begin{tikzpicture}
\raisebox{-1.3pt}{
\draw (0,0) circle (1ex);\fill (1ex,0) arc (0:#1:1ex) -- (0,0) -- cycle;
}
\end{tikzpicture}%
}

\usepackage{url}

\usepackage{booktabs}
\usepackage{threeparttable}

\usepackage{enumitem}

\begin{document}

\title{
A Blockchain-based Platform for Reliable\\
Inference and Training of Large-Scale Models
}

\pagestyle{plain}

\author{
    \IEEEauthorblockN{
    Sanghyeon Park
    }
    \IEEEauthorblockA{
        \textit{Seoul National University}\\
        Seoul, Republic of Korea\\
        \href{mailto:lukepark@snu.ac.kr}{lukepark@snu.ac.kr}
    }
    \and
    \IEEEauthorblockN{
    Junmo Lee
    }
    \IEEEauthorblockA{
        \textit{Seoul National University}\\
        Seoul, Republic of Korea\\
        \href{mailto:junmo.lee@snu.ac.kr}{junmo.lee@snu.ac.kr}
    }
    \and
    \IEEEauthorblockN{
    Soo-Mook Moon
    }
    \IEEEauthorblockA{
        \textit{Seoul National University}\\
        Seoul, Republic of Korea\\
        \href{mailto:smoon@snu.ac.kr}{smoon@snu.ac.kr}
    }
}

\maketitle

\begin{abstract}

As artificial intelligence (AI) continues to permeate various domains, concerns surrounding trust and transparency in AI-driven inference and training processes have emerged, particularly with respect to potential biases and traceability challenges.
Decentralized solutions such as blockchain have been proposed to tackle these issues, but they often struggle when dealing with large-scale models, leading to time-consuming inference and inefficient training verification.

To overcome these limitations, we introduce BRAIN, a Blockchain-based Reliable AI Network, a novel platform specifically designed to ensure reliable inference and training of large models.
BRAIN harnesses a unique two-phase transaction mechanism, allowing real-time processing via pipelining by separating request and response transactions.
Each randomly-selected inference committee commits and reveals the inference results, and upon reaching an agreement through a smart contract, then the requested operation is executed using the consensus result.
Additionally, BRAIN carries out training by employing a randomly-selected training committee. They submit commit and reveal transactions along with their respective scores, enabling local model aggregation based on the median value of the scores.

Experimental results demonstrate that BRAIN delivers considerably higher inference throughput at reasonable gas fees. In particular, BRAIN's tasks-per-second performance is 454.4293 times greater than that of a naïve single-phase implementation.

\end{abstract}



\section{Introduction} \label{section:introduction}

The rapid advancement of artificial intelligence (AI) based on large-scale deep artificial neural networks has transformed various industries by offering numerous powerful AI-based services \cite{brown2020language, Midjourney, agostinelli2023musiclm, rombach2022highresolution}. However, these networks typically depend on centralized servers for both learning and inference, raising concerns about trust, such as a lack of transparency in the learning process and the potential for manipulated inferences \cite{9347468, 8957108}.
For example, biases in AI algorithms can disproportionately affect certain racial and ethnic groups, as evidenced by issues in facial recognition systems~\cite{pmlr-v81-buolamwini18a} and criminal justice applications~\cite{brackey2019analysis}. The lack of transparency in the learning process makes it challenging for users to assess the fairness and ethical implications of these services, undermining trust in AI systems.
Similarly, centralized inference processes pose challenges due to potential manipulation, misuse, or difficulty in verifying the authenticity of AI-generated outputs. Examples include the proliferation of fake news, and the misappropriation of AI-generated inferences \cite{zhang2020overview, tolosana2020deepfakes}.

This lack of transparency can erode trust in AI services, as users struggle to determine the veracity. It can lead to increased costs for users, who may need to resort to separate searches, audits, or third-party evaluations to ensure trust.
Service providers may also face the burden of additional marketing efforts to convince users of their system's reliability.

\definecolor{Gray}{gray}{0.80}

\begin{table}
\begin{center}
\begin{adjustbox}{max width=1.0\columnwidth}
\begin{threeparttable}
    \caption{A comparison of BRAIN with existing studies}
    \begin{tabular}{l|c|c|c|c|c|c}
        \hline
            & \multicolumn{2}{c|}{Computational Engine}
            & \multicolumn{2}{c|}{Trust Machine}
            & \multicolumn{1}{c|}{Large}
            & \multicolumn{1}{c}{No} \\
            \cline{2-5} & Training & Inference & Training & Inference & Models & Hardfork \\
        \hline \hline
            DeepBrain \cite{17deepbrain} & \pie{360} & \pie{360} & \pie{0} & \pie{0} & \pie{360} & \texttimes{} \\
            Cortex \cite{chen2018cortex} & \pie{0} & \pie{360} & \pie{0} & \pie{360} & \pie{0}\tnote{\textdagger} & \texttimes{} \\
            AI News \cite{19aiNews} & \texttimes{} & \texttimes{} & \pie{0} & \pie{0} & \pie{360} & \pie{360} \\
            Flchain \cite{19flchain} & \pie{360} & \pie{0} & \pie{360} & \pie{0} & \pie{360} & \texttimes{} \\
            Baffle \cite{20baffle} & \pie{360} & \pie{0} & \pie{0} & \pie{0} & \pie{180}\tnote{\textdaggerdbl} & \pie{360} \\
            Fedcoin \cite{20fedcoin} & \pie{360} & \pie{0} & \pie{360} & \pie{0} & \pie{360} & \texttimes{} \\
            BFLC \cite{21blockDecFLCommittee} & \pie{360} & \pie{0} & \pie{360} & \pie{0} & \pie{360} & \texttimes{} \\
            Blockflow \cite{22blockflow} & \pie{360} & \pie{0} & \pie{360} & \pie{0} & \pie{360} & \pie{360} \\
            DFL \cite{tian2021dfl} & \pie{360} & \pie{0} & \pie{360} & \pie{0} & \pie{180}\tnote{\S} & \texttimes{} \\
            BlockDFL \cite{qin2023blockdfl} & \pie{360} & \pie{0} & \pie{360} & \pie{0} & \pie{360} & \texttimes{} \\
            iDML \cite{22blockflow} & \pie{360} & \pie{0} & \pie{360} & \pie{0} & \pie{360} & \pie{360} \\
        \hline
            ZEN \cite{feng2021zen} & \texttimes{} & \texttimes{} & \pie{0} & \pie{360} & \pie{0}\tnote{*} & \texttimes{} \\
            ZKP-FL \cite{xing2023zero} & \pie{360} & \pie{0} & \pie{360} & \pie{0} & \pie{0}\tnote{*} & \pie{360} \\
            VOC \cite{heiss2022advancing} & \pie{360} & \pie{0} & \pie{360} & \pie{0} & \pie{0}\tnote{*} & \pie{360} \\
            Y. Fan \cite{fan2023validating} & \texttimes{} & \texttimes{} & \pie{0} & \pie{360} & \pie{0}\tnote{*} & \texttimes{} \\
        \hline
            \rowcolor{Gray} BRAIN (Ours) & \pie{360} & \pie{360} & \pie{360} & \pie{360} & \pie{360} & \pie{360} \\
        \hline
    \end{tabular}
    \begin{tablenotes}
        \item[\textdagger] Negative impacts on transaction throughput.\\
        Moreover, it cannot support models whose inference time exceeds the block interval.
        \item[\textdaggerdbl] Due to the massive costs involved, on Ethereum~\cite{wood2014ethereum}.
        \item[\S] The cost depends on the model size.
        \item[*] Due to the tremendous time required.
    \end{tablenotes}
    \label{tab:feature-comparison}
\end{threeparttable}
\end{adjustbox}
\end{center}
\end{table}

\subsection{Related Work \& Challenges} \label{subsection:related}

The centralization of AI training and inference processes has been a significant issue, leading to the exploration of decentralized technologies as a solution. One such technology is blockchain.
However, integrating blockchain and AI has proven to be challenging. While some studies have made progress toward a solution, they have only partially succeeded.

\cite{19aiNews} have presented services that use blockchain and AI together but have only used blockchain as a database and incentive platform. They have not presented a way to increase trust in AI's training and inference directly.
\cite{20baffle} attempted to reduce aggregator costs through the blockchain but still failed to provide trust. Additionally, the high cost of storing model weights on the blockchain does not support large-scale neural networks effectively.
Meanwhile, other studies \cite{22blockflow, 20fedcoin, 19flchain} guided the training in the right direction through the use of incentives or slashing deposits, but did not provide confidence in inference.
\cite{21blockDecFLCommittee} presented a methodology for improving neural networks by evaluating and aggregating trained neural networks from committees, but it similarly did not consider inference.
Furthermore, in \cite{chen2018cortex}, although trust was established by enabling inference to be performed on the blockchain and allowing all participating nodes to validate inference results, it cannot support large neural networks that require a long time for inference due to the need for nodes to wait without processing other transactions. It also does not concern itself with training.
With \cite{17deepbrain}, participating nodes' GPUs can be rented and used for training and inference, but the trustworthiness of the results is not guaranteed.

Zero-knowledge proofs (ZKPs) can be used to verify the training and inference, as demonstrated in \cite{fan2023validating, xing2023zero, heiss2022advancing, feng2021zen}.
They do not require repeated verification; only proof is needed to verify that the operations were performed correctly.
However, because of the tremendous proof generation time, ZKPs are limited to small models, such as MobileNet~\cite{howard2017mob}.

A summary of these related studies can be found in Table~\ref{tab:feature-comparison}. We denote the level of support as \pie{360} for full support, \pie{180} for partial support, \pie{0} for no support, and \texttimes{} for not related features.
In particular, it can be observed that blockchain has not been fully utilized as a computational engine or a trust machine. Even when trust is provided, large models cannot be supported due to the time constraint or the high cost.

\subsection{Contributions} \label{subsection:contributions}

Our proposed solution, Blockchain-based Reliable AI Network (BRAIN), effectively addresses the challenges faced in previous studies.
To the best of our knowledge, BRAIN is the first blockchain platform capable of handling both inference and training for large neural networks while maintaining compatibility with existing blockchains.

The main contributions of BRAIN are as follows:

\begin{itemize}
    \item We propose BRAIN, an innovative architecture for large-scale neural networks, that addresses latency, redundancy, verification, and compatibility issues in existing chains.

    \item One of the key components of BRAIN is aggregator-free federated learning. This approach enables asynchronous model updates by utilizing smart contracts, leading to convergence without the need for an aggregator.

    \item Our simulation results demonstrate BRAIN's performance improvements and reasonable cost, and we provide guidelines for optimal hyperparameter configurations.

    \item We emphasize the importance of blockchain-powered AI for trust, transparency, and traceability. In addition, highlight BRAIN's potential impact on various applications.
\end{itemize}

Our implementation and experiment codes are publicly available for replication and further research at GitHub\footnote{\href{https://github.com/BRAIN-chain}{https://github.com/BRAIN-chain}}.

\section{Background} \label{section:background}

This section provides essential background related to the BRAIN protocol. We briefly discuss \textbf{Verifiable Random Functions}, which enable cryptographic sortition, and \textbf{Federated Learning}, a training approach with multiple participants.

\subsection{Verifiable Random Functions} \label{subsection:vrf}

Verifiable Random Functions (VRFs) \cite{micali1999verifiable} are functions that output verifiable pseudorandom values.
They consist of three functions: \texttt{keygen}, \texttt{evaluate}, and \texttt{verify}.

\begin{itemize}
    \item $\texttt{keygen}(r) \rightarrow (pk, sk)$: generates a public key $pk$ and a secret key $sk$ pair for a given random input $r$.
    \item $\texttt{evaluate}_{sk}(x) \rightarrow (y, \pi)$: returns pseudorandom output $y$ and a proof $\pi$ from the secret key $sk$ and $x$ as inputs.
    \item $\texttt{verify}_{pk}(x, y, \pi) \in \{\texttt{true}, \texttt{false}\}$: takes the public key $pk$, the pseudorandom output $y$, the proof $\pi$, and the message $x$ as inputs, and returns \texttt{true} if $y$ is actually the output produced by $sk$ and $x$. If not, it returns \texttt{false}.
\end{itemize}


Cryptographic Sortition is a method of selecting members based on cryptographic calculations, instead of predefined rules. Participants can confirm their own election but not others until the verifiable proofs are published.
In BRAIN, Cryptographic Sortition is implemented using VRF, similar to Algorand \cite{gilad2017algorand}. Since the $sk$ are not shared with one another, it is not possible to know who was elected to the committee for this round until the results are submitted. This makes it impossible to engage in malicious activities such as bribery.

\subsection{Federated Learning} \label{subsection:fl}

Federated learning is a technique that allows clients to train a global model by training local models on their own data. The aggregator then collects updates on the local models and integrates them to create the next global model. One well-known method for achieving this is Federated Averaging (FedAvg) \cite{mcmahan2017communication}, which aggregates updates from the clients through a weighted average based on the amount of data they hold. Furthermore, the FedAsync \cite{xie2019asynchronous} study demonstrates that learning can converge even in asynchronous environments.

BRAIN is designed to allow participants to perform model updates in parallel, inspired by the method of FedAsync.
However, BRAIN employs smart contracts instead of an aggregator for decentralization.
In addition, BRAIN utilizes a k-fold cross-validation technique similar to BFLC \cite{21blockDecFLCommittee} to evaluate update proposals. This technique uses local datasets held by each participant as a validation set, making the system resistant to attacks such as poisoning attacks~\cite{yang2017generative, biggio2012poisoning}.

\definecolor{Gray}{gray}{0.80}

\begin{figure}[!t]
    \centering
    \includegraphics[width=1.0\linewidth]{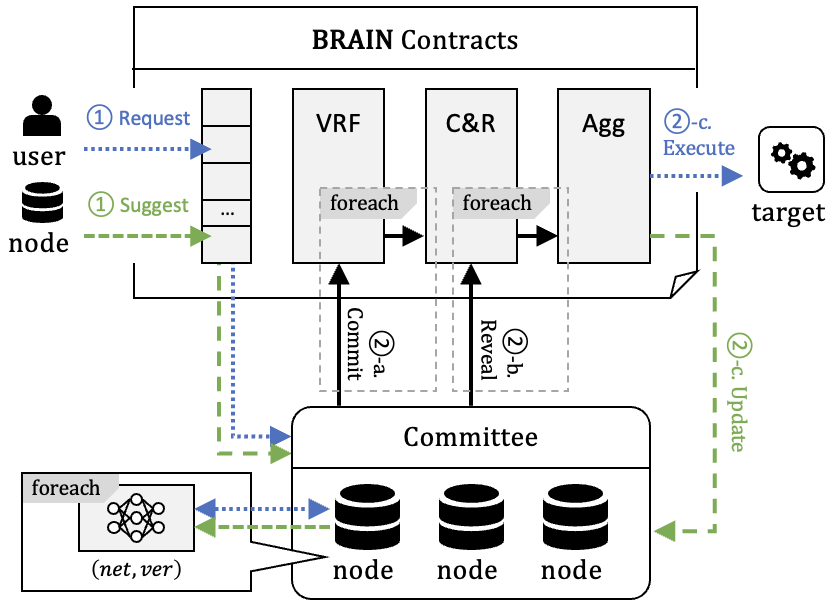}
    \caption{
        The figure illustrates the unified process flow for both inference (dotted blue line) and training (dash-dotted green line) in the BRAIN platform.
        \textcircled{\raisebox{-0.9pt}{1}} A participant submits a request for either inference or model update evaluation, which is then added to the queue.
        \textcircled{\raisebox{-0.9pt}{2}}-a. Each member of the respective committee (inference or training) submits a commitment.
        \textcircled{\raisebox{-0.9pt}{2}}-b. Each committee member reveals their submitted value (inference output or model score).
        \textcircled{\raisebox{-0.9pt}{2}}-c. In the case of inference, anyone can execute the requested action using the inference output. In the case of training, BRAIN nodes obtain the median value of evaluation scores and use that value to compute the updated model locally.
    }
    \label{fig:overview}
\end{figure}

\section{BRAIN: Overview} \label{section:brain-design}

BRAIN is a decentralized, blockchain-based AI platform that provides both training and inference services.
This section presents a comprehensive overview of BRAIN, focusing on the roles of \textbf{participants} and the significance of \textbf{hyperparameters} in maintaining the platform's optimal operation.

\subsection{Types of Participants} \label{subsection:participants}

BRAIN's ecosystem consists of various types of participants, each of which can assume multiple roles:

\vspace{4pt}

\noindent \textbf{Users.} Users are individuals who access the inference services provided by BRAIN. They spend assets, such as ETH, to submit inference requests through transactions and instruct how the resulting outputs should be processed.
In Fig.~\ref{fig:overview}, \textcircled{\raisebox{-0.9pt}{1}} \textsc{Request} represents the initiation of an inference request.

\vspace{4pt}

\noindent \textbf{BRAIN Nodes.} Nodes are participants who have deposited sufficient assets, enabling them to contribute computational resources for inference, training, or both within the BRAIN ecosystem.
Depositing assets allows nodes to actively participate in the system, earning fees from users while also risking having their deposits slashed through smart contracts in the event of malicious behavior.
In this paper, it is assumed that all nodes have deposited the same amount of assets. This assumption is made without loss of generality, as the BRAIN does not limit the number of participants, allowing for flexibility in the distribution of deposits into multiple nodes.
A BRAIN node may belong to an Inference Committee, a Training Committee, both committees, or neither.

\vspace{4pt}

\noindent \textbf{Inference Committee.} The Inference committee is composed of BRAIN nodes selected through a VRF and is responsible for handling inference requests. A different configuration of nodes is selected for each request. Committee members receive incentives for participating in the processing of inference.

\vspace{4pt}

\noindent \textbf{Training Committee.} The Training committee, similarly consisting of nodes chosen through a VRF, is responsible for evaluating model update proposals. They receive incentives for their contributions to the evaluation of model updates.

\vspace{4pt}

To prevent free-riders who do not perform tasks but merely copy another entity's result, BRAIN adopts a Commit-and-Reveal scheme, denoted as \textcircled{\raisebox{-0.9pt}{2}}-a \textsc{Commit} in Fig.~\ref{fig:overview}, followed by \textcircled{\raisebox{-0.9pt}{2}}-b \textsc{Reveal}.
After the revelation, the agreed-upon result is used for the AI service's response as \textcircled{\raisebox{-0.9pt}{2}}-c \textsc{Execute} or to update the model parameters as \textcircled{\raisebox{-0.9pt}{2}}-c \textsc{Update}.

\vspace{4pt}

If participants engage in malicious or lazy behavior that harms the protocol, they will be punished to deter such negative behavior. In BRAIN, punishment is carried out by slashing assets deposited.
Users who are non-deposit entities may also engage in malicious behavior, such as launching Denial-of-Service attacks with numerous meaningless inference requests \cite{saad2019mempool, saad2018poster}. However, these attempts can be thwarted by making attackers spend massive assets, similar to how transaction fees are used in blockchain systems.
Furthermore, preventing malicious behavior by block producers is the responsibility of the underlying blockchain protocol, not BRAIN.

\subsection{Hyperparameters} \label{subsection:hyperparameters}

In BRAIN, predefined values known as hyperparameters play a crucial role in ensuring the platform's optimal operation.

\begin{itemize}
    \item{$E$ and $d$:} $E$ denotes the epoch at which the VRF input changes, while $d$ adjusts the probability of committee selection.
    $E$ should be set appropriately to prevent VRF verification failure or slow quorum formation. Adjusting $d$ can minimize the number of selected nodes, reducing costs but trading off security from redundancy.
    Inference and training have their own epochs ($E_I$ and $E_T$, respectively) and probabilities ($d_I$ and $d_T$, respectively).

    \item{$f$:} The finality factor ensures the VRF \textit{seed}'s unchangeability from chain forks. The unit of $f$ is round, which increases by $1$ each time the \textcircled{\raisebox{-0.9pt}{2}}-a phase is completed.

    \item{$Q_C$ and $Q_R$:} They determine the number of nodes needed for quorum in \textcircled{\raisebox{-0.9pt}{2}}-a and \textcircled{\raisebox{-0.9pt}{2}}-b respectively. Adjusting them in relation to $d$ can balance liveliness and performance.

    \item{$T_C$, $T_R$, $T_E$, and $T_U$:} Timeout values control various aspects of BRAIN. $T_C$ in \textcircled{\raisebox{-0.9pt}{2}}-a should be set to cancel unwanted requests, such as computationally-intensive requests. While $T_E$ and $T_U$ in \textcircled{\raisebox{-0.9pt}{2}}-c may be set to infinity since each request has its own $timeout$ field. $T_R$ in \textcircled{\raisebox{-0.9pt}{2}}-b is used to penalize nodes that don't reveal on time.

    \item{$R_C$, $R_E$, $R_U$, $R_R$, and $R_S$:}
    These symbols represent rewards for successfully completing \textsc{Commit}, \textsc{Execute}, \textsc{Update}, \textsc{Reveal}, and \textsc{Suggest} tasks, respectively. Rewards incentivize nodes to perform their assigned tasks.
    Essentially, the funding for this is provided through the user's inference request fee.
    It is recommended to set $R_C$ to $0$ to encourage nodes to submit the following \textsc{Reveal}.

    \item{$P_C$, $P_R$, and $P_S$:}
    Specifically, $P_C$ is the penalty for a faulty \textsc{Commit}, $P_R$ is for an invalid or no \textsc{Reveal}, and $P_S$ is for a faulty \textsc{Suggest}. 
    Penalties prevent attacks and encourage honest inference and evaluation. They should be set appropriately to promote positive resource usage.
\end{itemize}

We use these notations throughout the paper. In particular, in section~\ref{section:experiment}, we discuss and examine how hyperparameters affect BRAIN's performance and the stability of its services.

\subsection{Design Goal \& Features} \label{subsection:goal-challenges}

As described in section~\ref{subsection:related}, integrating AI and blockchain poses various challenges, particularly when dealing with large models.
We have designed several features to address the corresponding challenges, which are briefly discussed below:

\vspace{4pt}

\noindent \textbf{Minimized Latency with Transaction Pipelining.} We design a system that minimizes the latency associated with AI, to mitigate negatively impacting the overall performance of the blockchain. To achieve this, we have implemented a two-phase transaction process, represented as \textcircled{\raisebox{-0.9pt}{1}} and \textcircled{\raisebox{-0.9pt}{2}} in Fig.~\ref{fig:overview}, that separates the request and response transactions. 
Furthermore, the \textcircled{\raisebox{-0.9pt}{2}} phase is further divided into three subphases: \textcircled{\raisebox{-0.9pt}{2}}-a, \textcircled{\raisebox{-0.9pt}{2}}-b, and \textcircled{\raisebox{-0.9pt}{2}}-c.
This approach allows pipelining for efficient real-time processing, reducing the strain on the blockchain.

\vspace{4pt}

\noindent \textbf{Scalable Verification with Cryptographic Sortition.} We implement a scalable verification mechanism that balances the need for trust with the computational costs of AI inference and training. To achieve this, BRAIN utilizes cryptographic sortition based on verifiable random functions to select a random subset of nodes, called the committee, to perform the inference or training process. This approach enhances verification efficiency by eliminating redundancy while ensuring that a centralized entity does not manipulate the process.

\vspace{4pt}

\noindent \textbf{Decentralized Training with Aggregator-Free Federated Learning.} We aim to establish a decentralized training process capable of handling large-scale models and preventing manipulation. To achieve this, BRAIN employs aggregator-free federated learning. A randomly selected committee of nodes, the training committee, evaluates and reaches a consensus on the score for newly submitted model updates. These agreed-upon scores are saved in the smart contract and used as weights to calculate the integrated model. Each node performs this calculation locally without the need for a centralized aggregator. This approach enables decentralized and efficient learning that can withstand the presence of malicious nodes.

\vspace{4pt}

\noindent \textbf{Compatibility with Existing Blockchains.} BRAIN is designed to be compatible with existing contract-aware blockchains, eliminating the need for hard forks and enabling smooth integration with existing high-secure networks.

\newcommand{\RNum}[1]{\uppercase\expandafter{\romannumeral #1\relax}}

\section{BRAIN: Inference} \label{section:brain-inference}

The inference component of BRAIN is designed to provide trust and security in a decentralized environment. It has several features, including a \textbf{two-phase transaction} structure, \textbf{cryptographic sortitions}, and a \textbf{commit-and-reveal} scheme.

\subsection{Two-Phase Transaction} \label{subsection:two-phase}

\begin{figure*}[!t]
    \centering
    \includegraphics[width=1.0\linewidth]{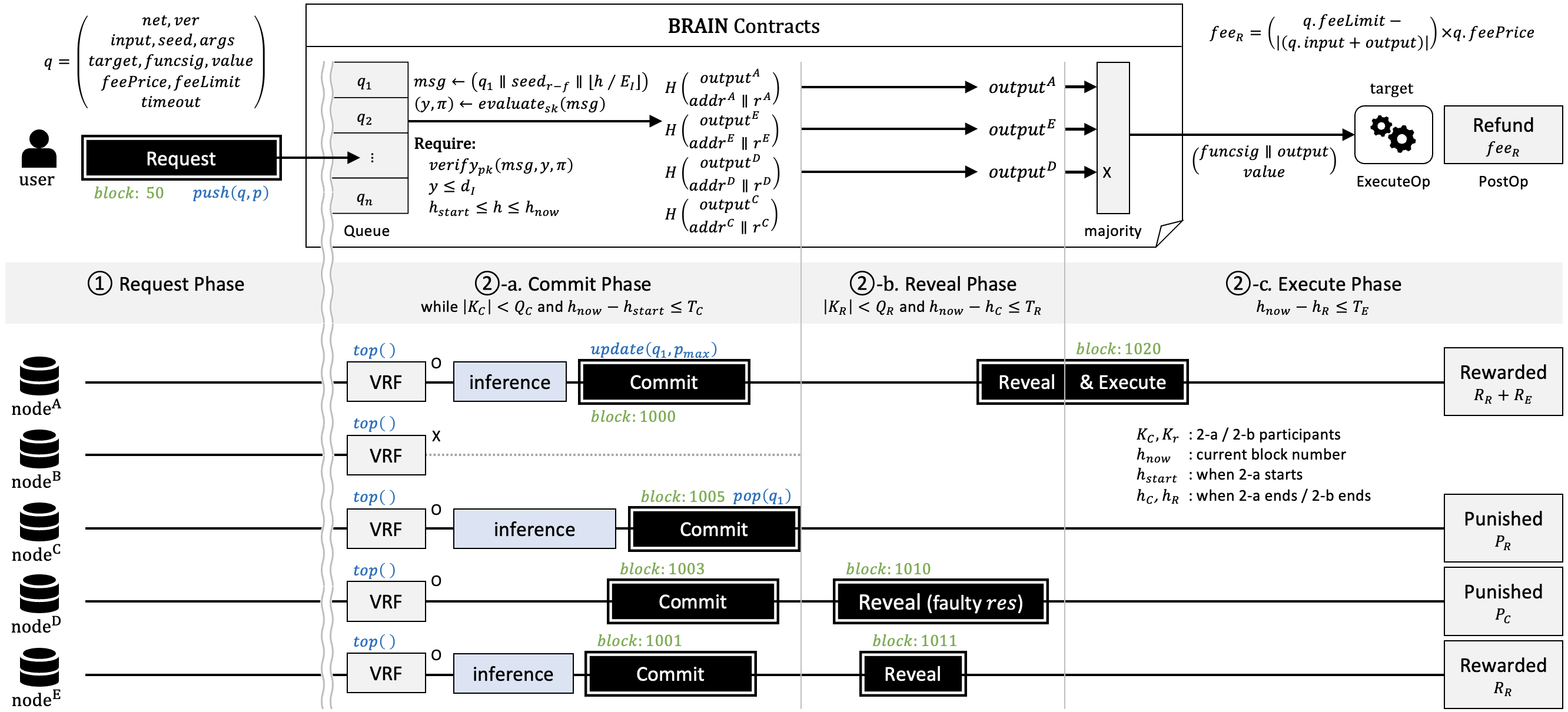}
    \caption{Overview of the process of verifying inference in BRAIN. A user sends an inference request in phase \textcircled{\raisebox{-0.9pt}{1}}. When the request is popped from the priority queue, phase \textcircled{\raisebox{-0.9pt}{2}} begins. The results are then committed in phase \textcircled{\raisebox{-0.9pt}{2}}-a through an inference committee. In phase \textcircled{\raisebox{-0.9pt}{2}}-b, the original results are revealed. If a quorum is reached, the final result is obtained through a majority in phase \textcircled{\raisebox{-0.9pt}{2}}-c and the requested operation is executed. The reward payment and punishment are made in the \textsc{PostOp} step. In addition, the contract refunds the remaining fee to the user. In the figure, a black square with a double border represents a transaction that is publicly recorded on the chain. The green block numbers indicate the example order of events in the process.}
    \label{fig:inference}
\end{figure*}

One way to enable inference on a blockchain is to add \textsc{inference} operations as commands, as in \cite{chen2018cortex}. This allows for direct invocation of inference on the chain, but the real-time nature of complex neural network inference may not be feasible due to slower processing speeds compared to ordinary transactions and even block generation times.

The BRAIN protocol addresses the issue of delays in processing AI tasks on the blockchain by separating the request and response into two distinct transactions. The first phase involves pushing the inference request into a queue, while the second phase involves popping it from the queue, performing the inference, and disclosing the result in a separate transaction. This two-phase configuration allows the inference request transaction to be processed immediately, enabling block producers to move on to the next transaction in the block without waiting for a response.
We use a priority queue implemented through smart contracts to hold and process request transactions. Requests are added to the queue and processed in order of priority, allowing users to prioritize their requests by paying for more assets. This is similar to the concept of transaction fees (gas) in Ethereum \cite{wood2014ethereum}.

\subsection{Cryptographic Sortition} \label{subsection:cryptography-sortition}

Cryptographic Sortition is a key feature of the BRAIN protocol that ensures both effectiveness and trustworthiness of the inference committee.
It reduces costs and improves network performance by limiting the number of nodes that perform redundant operations.
An attacker seeking to produce false inference results would need to occupy a majority of the inference committee and take control of the entire process of phase \textcircled{\raisebox{-0.9pt}{2}}. However, this is made difficult as the identities of the selected committee members are kept secret.

Since BRAIN nodes are already participants in the blockchain system, they possess their own $pk$ and $sk$ keys. Hence, these keys can be used in the VRF process instead of generating new ones using the \texttt{keygen} function.
As shown in Fig.~\ref{fig:inference}, $msg$, the input to the VRF functions \texttt{verify} and \texttt{evaluate}, consists of several elements used to ensure the reliability and stability of the inference results.
The highest-priority inference request $q_1$ is included to indicate that it corresponds to the correct request.
The random value ${seed}_{r-f}$, stored in the contract, introduces randomness into the message to reduce the chances of manipulating the process, like \cite{gilad2017algorand}.
Using the $f$-th past seed rather than the latest ${seed}_r$ minimizes the risk of the $msg$ changing due to forks.
$\lfloor h / E_I \rfloor$ increases the diversity of $msg$ according to pre-defined hyperparameter $E_I$, which adds additional committee members to ensure the liveness of the inference process.

There are two ways to calculate VRF on the contract:
1) Implementing the whole verification logic through contracts. However, this method incurs high costs due to the complexity of the elliptic curve operations.
2) Using the precompiled-contract \texttt{ecrecover} that is already built into the EVM. It significantly reduces costs compared with the first method. The security level is decreased from 32 bytes (hash) to 20 bytes (address) when using this method, while this is generally sufficient.
Section~\ref{subsection:contract-overhead} provides a comprehensive analysis of the costs associated with both methods.

\subsection{Commit-and-Reveal} \label{subsection:commit-and-reveal}

The commit-and-reveal scheme during the phase \textcircled{\raisebox{-0.9pt}{2}} is designed to prevent free-riding. Without this scheme, nodes could potentially view the others' inference results in transactions and submit the transaction with the same value without actually performing the required inference operations. The commit-and-reveal scheme in BRAIN ensures that the result with a random value remains hidden, so no information can be obtained by viewing the hash value. Additionally, the hash input includes an address, essentially making it a commit that other nodes cannot use. Therefore, the nodes of the committee must perform the required operations and prove they have committed the correct result during the subsequent reveal phase, thus holding them accountable for their actions.

\subsection{Inference Process} \label{subsection:inference-process}

\noindent \textbf{Phase \textcircled{\raisebox{-0.9pt}{1}}.}
As in Fig.~\ref{fig:inference}, phase \textcircled{\raisebox{-0.9pt}{1}} of the two-phase transaction pushes an inference request $q$ into the priority queue.
The request includes a unique identifier $(net, ver)$ for what neural network model is requested.
Along with $input$, we use a random $seed$ to address the non-deterministic features of the model. Other AI-related parameters, such as {\em max\_length} and {\em temperature}, are likewise included in the request as serialized bytes in $args$ to coordinate the neural network.
Furthermore, the request specifies the address $target$, the function $funcsig$, and the native asset amount $value$ to be called with the inference result. This request is stored in the queue for later processing by the committee. 
$timeout$ is specified to prevent requests from being processed too late.

The fee for the request is typically calculated based on the length of both the $input$ and $output$, similar to the OpenAI pricing~\cite{openai}. However, as the length of the $output$ cannot be determined in advance, users are required to pay sufficient fees upfront.
To achieve this, the system employs a fee pricing mechanism using the parameters $feePrice$ and $feeLimit$, which is similar to the gas system used by Ethereum before EIP-1559~\cite{eip1559}.
The request is canceled if the combined length of the $input$ and $output$ exceeds the specified $feeLimit$. The total amount paid by the user in advance is determined by multiplying $feeLimit$ by $feePrice$, but the actual amount paid is calculated as $(input + output) \times feePrice$, and the remainings are refunded in \textcircled{\raisebox{-0.9pt}{2}}-c phase.
The priority of the request is determined by the $feePrice$.

\vspace{4pt}

\noindent \textbf{Phase \textcircled{\raisebox{-0.9pt}{2}}.}
BRAIN is designed to handle inference requests securely. It does so through a VRF-based cryptographic sortition process that elects inference committees and a commit-and-reveal scheme that registers values and reveals them later.
As illustrated in Fig.~\ref{fig:inference}, the \textsc{Commit} process is divided into \textcircled{\raisebox{-0.9pt}{2}}-a phases, while the \textsc{Reveal} process is divided into \textcircled{\raisebox{-0.9pt}{2}}-b phases. In addition, the steps of processing a user's request with the inference $output$ after the revelation are divided into \textcircled{\raisebox{-0.9pt}{2}}-c \textsc{Execute} phases, where execution is combined with the last revelation into a single transaction by the contract.

\vspace{4pt}

\noindent \textbf{Phase \textcircled{\raisebox{-0.9pt}{2}}-a.} Nodes self-verify their election for the inference committee for a particular inference request, using VRF. If the result of \texttt{evaluate}, 256-bit value $y$, is less than or equal to the difficulty condition hyperparameter $d_I$, the node is eligible to participate as an inference committee member. The nodes that pass the \texttt{verify} by the contract are eligible to register the inference result as a member of the inference committee. The hashed result $output$, along with the address $addr$ and a random value $r$, is registered in the contract through a \textsc{Commit} transaction.
To proceed from \textcircled{\raisebox{-0.9pt}{2}}-a to \textcircled{\raisebox{-0.9pt}{2}}-b, more than $Q_C$ \textsc{Commit} transactions must be published.

\vspace{4pt}

\noindent \textbf{Phase \textcircled{\raisebox{-0.9pt}{2}}-b.} The process of revealing the previously committed value is performed. To verify the revealed value matches the previously registered value in the contract, the same components used to construct the hash value — $output$, $addr$, and $r$ — are passed to the contract. The contract calculates the hash again using these values, and if the revealed value differs from the committed value, the \textsc{Reveal} transaction is rejected.
When nodes with a quorum of at least $Q_R$ reveal their committed values, the process can proceed to phase \textcircled{\raisebox{-0.9pt}{2}}-c.

\vspace{4pt}

\noindent \textbf{Phase \textcircled{\raisebox{-0.9pt}{2}}-c.}
Since BRAIN does not have a trusted subject, the final result is derived through a majority of the revealed values.
Then function $funcsig$ requested by the user has executed on the target address $target$ with the $value$ and the agreed-upon result of the inference $output$ through the \textsc{ExecuteOp} step in \textsc{Execute} transaction.
In general, the results of all nodes should be consistent because of fixed $seed$. If the majority of the inference committee is honest, the real answer can always be agreed upon, and a node that reveals a result that violates this can be deemed to have taken malicious action and be penalized through the \textsc{PostOp} step. In addition, at the same step, the remaining fees are refunded to the user.

\vspace{4pt}

Thanks to the separation of subphases \textbf{a}, \textbf{b}, and \textbf{c} in phase \textcircled{\raisebox{-0.9pt}{2}}, inference processing can be pipelined. Phase \textcircled{\raisebox{-0.9pt}{2}}-b and \textcircled{\raisebox{-0.9pt}{2}}-c of the highest priority inference request $q_1$ can be performed concurrently with phase \textcircled{\raisebox{-0.9pt}{2}}-a of the next priority $q_2$, enabling a low-latency service.
Therefore, the throughput of inference transactions is only limited by the $Q_C$, $d$, and $T_C$, which determine the processing speed of the \textcircled{\raisebox{-0.9pt}{2}}-a phase.

\section{BRAIN: Training} \label{section:training}

BRAIN employs federated learning, in which the proposed model is evaluated by a training committee using the \textbf{model scoring}, and the update proposal is then aggregated through the \textbf{Aggregator-Free Federated Learning} methodology to derive a global model. This method enable anyone to propose updates at any time, and since they are validated and then aggregated, the model can be continuously improved.

\subsection{Model Scoring} \label{subsection:model-scoring}

A straightforward method for training verification would be to have the elected committee nodes perform the same learning process using fixed seeds and the same data. While this approach can ensure the reliability of model updates, it is inefficient due to the high cost of training, as a large number of nodes would perform redundant operations.

To address these issues, BRAIN employs techniques found in previous studies \cite{22blockflow, 21blockDecFLCommittee}. Each node reports a score after evaluating the model on its own dataset, similar to K-fold cross-validation \cite{21blockDecFLCommittee}.
To ensure fairness in scoring, the node that proposed the update is not allowed to participate in the committee, even if it has been qualified through the VRF. The score is typically based on accuracy. The median value is used to reach a consensus on the global score, as this method can provide a reliable result unless the majority of nodes are malicious. The model scoring method of BRAIN simplifies the role of the training committee in evaluating the model's performance rather than verifying the training process.

\subsection{Aggregator-Free Federated Learning} \label{subsection:aafl}

In traditional federated learning, a central aggregator computes and shares a global model that is the weighted average of updates. However, BRAIN has no central aggregator and no servers. Instead, each node can locally aggregate the model using the Weighted Moving Average (WMA) from the stored scores at the contract.
Although \cite{20baffle} uses contracts for weight storage, it is limited by massive transaction fees and cannot support large models, as discussed in section~\ref{subsection:related}.
BRAIN can significantly reduce upload costs regardless of the model size by storing only the model's score and not the model itself. This approach is similar to the concept of reputation used by \cite{tian2021dfl}, as well as the voting mechanisms employed by \cite{qin2023blockdfl, yu2023idml}.

\begin{algorithm}[H]
\caption{Aggregator-Free Federated Learning}
\begin{algorithmic}[1]

\Require \text{window size} $n \ge 2$, \text{threshold} ${score}_{th} \le a_r$
\Ensure \text{locally calculated global model} $\overline{M}$


\Procedure{AaflWma}{$n$}
    \State Initialize $r \gets 0, \overline{M_0} \gets M_0, a_0 = 1$
    
    \ForAll{\text{new} $(M_r, a_r)$ and $r \ge 1$}
        \State $\overline{M_r} \gets (1-\alpha) \overline{M_{r-1}} + \alpha M_r$, 
        \State \textit{where} \[
            \alpha = \begin{cases}
                \frac{a_r}{a_{r-n+1}+...+a_r} & \text{if $r \ge n-1$}\\
                \frac{a_r}{a_0 + ... + a_r} & \text{otherwise}
            \end{cases} \]
    \EndFor
\EndProcedure

\end{algorithmic} 
\label{algo:affl-wma}
\end{algorithm}

As described in algorithm~\ref{algo:affl-wma}, the initial global model $\overline{M_0}$ is set equal to $M_0$. For each $r$-th global model $\overline{M_r}$, nodes calculate the WMA of the previous $n$ models, including the current update suggestion $M_r$, using corresponding scores $a_{r-n+1}, ..., a_r$ as weights. Since all nodes start from the same model and integrate updates with the same scores, locally derived global models will be consistent with each other, unless a node intentionally omits updates or misuses weights.

The convergence of this technique can be proven by substituting the WMA equation into the form of FedAsync~\cite{xie2019asynchronous}, which has been mathematically proven to converge. 

\begin{proof}
Because $\overline{M_{r-1}}=\frac{a_{r-n+1}M_{r-n+1}+...+a_{r-1}M_{r-1}}{a_{r-n+1}+...+a_{r-1}}$,
\begin{equation*}
\begin{split}
    \overline{M_r} &= \frac{(a_{r-n+1}M_{r-n+1} + ... + a_{r-1}M_{r-1} + a_rM_r)}{(a_{r-n+1} + ... a_r)}\\
    &= (1-\frac{a_r}{a_{r-n+1} + ... a_r})\overline{M_{r-1}} + (\frac{a_r}{a_{r-n+1} + ... a_r})M_r
\end{split}
\end{equation*}
Taking $\alpha = \frac{a_r}{a_{r-n+1}+...+a_r}$, we have
\begin{equation*}
\overline{M_r} = (1-\alpha) \overline{M_{r-1}} + \alpha M_r
\end{equation*}
which its convergence has been proven by \cite{xie2019asynchronous}.
Similar to $\alpha = \frac{a_r}{a_0 + ... + a_r}$ where $r < n-1$, without loss of generality.
\end{proof}

\subsection{Training Process} \label{subsection:training-process}

\noindent \textbf{Phase \textcircled{\raisebox{-0.9pt}{1}}.} As depicted in Fig.~\ref{fig:overview}, during the first phase of the training process, model update $M$ is published to the blockchain by \textsc{Suggest} transaction and pushed to a FIFO queue for processing. A circular queue is utilized to improve storage-cost efficiency, with details provided in section~\ref{subsection:contract-overhead}.
The proposed transaction includes an identifier $(net, ver)$, which specifies the target updated neural network with the version, and a $timeout$ field that indicates whether the update is outdated or not.
The $ver$ value is calculated as a hash of the neural network weights. Model updates are shared through any network such as IPFS \cite{benet2014ipfs} rather than the blockchain, allowing each node to validate the shared model by obtaining a hash of the weights and comparing it to the $ver$ stored in the contract.

\vspace{4pt}

\noindent \textbf{Phase \textcircled{\raisebox{-0.9pt}{2}}.} During the second phase of the training process in BRAIN, the model update proposal stored in the queue is processed by the training committee. This phase is similar to the inference process, with a few key differences. First, values that are hidden and revealed during phases \textcircled{\raisebox{-0.9pt}{2}}-a and \textcircled{\raisebox{-0.9pt}{2}}-b are $score$s rather than inference $output$s. Second, the median value is used instead of a majority during phase \textcircled{\raisebox{-0.9pt}{2}}-c. In addition, there is no need for an \textsc{ExecuteOp} step in the training process. During the \textsc{PostOp} step, rewards $R_U$ are given for the executing update. If the agreed-upon $score$ exceeds a certain threshold, the node that proposed the model update is rewarded with $R_S$. On the other hand, if the threshold is not met, the node is penalized with $P_S$.

\vspace{4pt}

Similar to the inference process, evaluation in the training process can be pipelined. This allows for efficient updates, as subphase \textcircled{\raisebox{-0.9pt}{2}}-b of the first update suggestion $M_1$ and phase \textcircled{\raisebox{-0.9pt}{2}}-a of the second suggestion $M_2$ can be started simultaneously.

\section{BRAIN: Implementation Details} \label{section:implementation}

In this section, we provide an overview of the implementation details for the BRAIN platform, focusing on the fallback mechanisms employed during the inference process.
The platform is designed to handle various scenarios where the quorum is not met or round timeouts occur, thus providing a robust and realtime decentralized AI platform.

In phase \textcircled{\raisebox{-0.9pt}{2}}-a, if a block above the timeout $T_C$ passes without the quorum $Q_C$ being met, the process falls into the \textbf{a-fallback} of algorithm~\ref{algo:inference-phase-2-a}, and the inference request $q$ is canceled and all assets are returned to the user.
The last \textsc{Commit} transaction that satisfies the quorum and meets the necessary requirements performs a $pop$ operation on the queue, ending phase \textcircled{\raisebox{-0.9pt}{2}}-a and beginning phase \textcircled{\raisebox{-0.9pt}{2}}-b.

If no response is received from nodes at least quorum $Q_R$ until the timeout $T_R$, a \textbf{b-fallback} is performed. As shown in the algorithm~\ref{algo:inference-phase-2-b}, depending on the trade-off between safety and liveness the service intends to provide, one of the following two methods is defined in advance:

\begin{enumerate}
    \item {b-{\RNum{1}}-fallback.} Although the quorum was not reached, only the revealed values proceeded to phase \textcircled{\raisebox{-0.9pt}{2}}-c. The degree of safety may decrease, but the service's liveness increase.

    \item {b-{\RNum{2}}-fallback.} The process ends here. The inference request $q$ is set to have the secondary-highest priority $p_{max}-1$ and is pushed back into the priority queue so that another set of inference committees can immediately process it. In this case, safety can be provided at a high level, but the degree of liveness decreases.
\end{enumerate}


\begin{algorithm}[H]
\caption{\textcircled{\raisebox{-0.9pt}{2}}-a. Commit Phase Pseudocode}
\begin{algorithmic}[1]

\renewcommand{\algorithmiccomment}[1]{{\color{blue}{\hfill$\triangleright$ #1}}}

\Require $top() \ne \textit{null}$
\Procedure{ResultCommit}{$Q_C$, $T_C$, $d_I$}
\State $q_1 \gets top()$
\For{{block height} $h \gets h_{start}, ...$}
    \For{$k \in \textit{\{nodes\}}-K_C$ {in parallel}}
        \State ${msg} \gets (q_1 \parallel {seed}_{r-f} \parallel \lfloor {h / E_I} \rfloor)$
        \State $(y, \pi) \gets \texttt{evaluate}_{sk}(msg)$
        \If{$y \le d_I$}
            \State ${output}^k \gets$ \textsc{Inference($q_1$)}
            \State $H \gets H({output}^k \parallel addr^k \parallel r^k)$
            \State \textbf{call} \textsc{Commit}$_{q_1}^k$($msg$, y, $\pi$, $H$)
        \EndIf
    \EndFor
\State \textbf{if} {$|K_C| \ge Q_C$} \textbf{then break} \Comment{End of \textcircled{\raisebox{-0.9pt}{2}}-a}
\State \textbf{else if} {$h - h_{start} > T_C$} \textbf{then break} \Comment{a-fallback}
\EndFor
\EndProcedure


\Function{Inference}{$q$}
    \State \Return ${q.net}_{q.ver}(q.input, q.seed, q.args)$
\EndFunction


\algrenewcommand\algorithmicprocedure{\textbf{transaction}}
\renewcommand{\algorithmiccomment}[1]{{\color{blue}{\hfill$\triangleright$ #1}}}

\Require $\texttt{verify}_{pk}(msg, y, \pi)$, $y \le d_I$, $h_{start} \le h \le h_{now}$
\Procedure{Commit$_q^k$}{$y$, $\pi$, $h$, $H$}
\State Store $H^k_q$ to BRAIN contract; $K_C \gets K_C \cup \{k\}$
\State \textbf{if} {$|K_C| == 1$} \textbf{then} $update(q, p_{max})$
\If{$|K_C| \ge Q_C$} \Comment{End of \textcircled{\raisebox{-0.9pt}{2}}-a}
    \State $pop(q)$; $h_{C} \gets h_{now}$
    \State $(seed_r, \pi) \leftarrow \texttt{evaluate}_{sk}(seed_{r-1} \parallel r)$
\ElsIf{$h_{now} - h_{start} > T_C$} \Comment{a-fallback}
    \State $pop(q)$; $K_C \gets \varnothing$
    \State $seed_r \leftarrow H(seed_{r-1} \parallel r)$
    \State Refund $({q}.feeLimit - \left| q.input \right|) \times q.feePrice$
    \State Refund ${q}.value$ \texttt{ETH}
\EndIf
\EndProcedure



\algstore{bkbreak}

\end{algorithmic}
\label{algo:inference-phase-2-a}
\end{algorithm}


\begin{algorithm}[H]
\caption{\textcircled{\raisebox{-0.9pt}{2}}-b. Reveal Phase Pseudocode}
\begin{algorithmic}[1]

\algrestore{bkbreak}

\renewcommand{\algorithmiccomment}[1]{{\color{blue}{\hfill$\triangleright$ #1}}}

\Require $K_C \ne \varnothing$, phase \textcircled{\raisebox{-0.9pt}{2}}-b fallback type $\tau \in \{ 
        \tau_{\RNum{1}},
        \tau_{\RNum{2}} 
    \}$
\Procedure{ResultReveal$_q$}{$Q_R$, $T_R$, $\tau$}
\For{{block height} $h \gets h_{C}, ...$}
    \For{$k \in K_C - K_R$ {in parallel}}
        \State \textbf{call} \textsc{Reveal}$_q^k$(${output^{k}}$, ${addr^{k}}$, ${r^{k}}$)
    \EndFor


    \State \textbf{if} {$|K_R| \ge Q_R$} \textbf{then break} \Comment{End of \textcircled{\raisebox{-0.9pt}{2}}-b}
    \State \textbf{else if} {$h - h_{C} > T_R$} \textbf{then break} \Comment{b-fallback}
\EndFor
\EndProcedure


\algrenewcommand\algorithmicprocedure{\textbf{transaction}}
\renewcommand{\algorithmiccomment}[1]{{\color{blue}{\hfill$\triangleright$ #1}}}

\Require $H(output \parallel addr \parallel r) == H^k_q$, $addr == {addr}^k$
\Procedure{Reveal$_q^k$}{$output$, $addr$, $r$}
    \State Store $output$ to BRAIN contract; $K_R \gets K_R \cup \{k\}$
    \If {$|K_R| \ge Q_R$} \textbf{call} \textsc{Execute} \Comment{End of \textcircled{\raisebox{-0.9pt}{2}}-b}
    \ElsIf {$h_{now} - h_{C} > T_R$}
        \State \textbf{if} $\tau == \tau_{\RNum{1}}$ \textbf{then} $h_{R} \gets h_{now}$ \Comment{b-{\RNum{1}}-fallback}
        \State \textbf{else} $push(q, p_{max}-1)$; $K_C \gets \varnothing$; $K_R \gets \varnothing$
        \Statex \Comment{b-{\RNum{2}}-fallback}
    \EndIf

\EndProcedure

    
\end{algorithmic}
\label{algo:inference-phase-2-b}
\end{algorithm}



\pgfplotsset{
    params/.style={
        fill=blue!30!white,
        draw=blue,
    },
    size/.style={
        fill=red!30!white,
        draw=red,
    },
    nodes near coords style={
        /pgf/number format/.cd,
            fixed,
            fixed zerofill,
            precision=2,
            1000 sep={},
        /tikz/.cd
    },
}

\newcommand{\myfontsize}[1]{\fontsize{#1}{#1}\selectfont}

\section{Experiments} \label{section:experiment}

To evaluate the performance and effectiveness of the BRAIN, we conducted a series of experiments: \textbf{inference transaction throughput} and \textbf{contract overhead analysis}. The results of these experiments provide insight into the performance and scalability of the BRAIN and demonstrate its ability to handle large neural networks. 


\subsection{Inference Transaction Throughput} \label{subsection:inference-troughput}

In this section, we evaluate the performance of the BRAIN platform in providing on-chain inference, focusing on metrics such as tasks-per-second and the number of timeouts.

The key variables examined in this experiment are:
We investigate the frequency of request transactions $freq$, with a default value of $0.0577 = (5752729 / 99638953)$, derived from the ratio of OpenSea~\cite{opensea} — the most transaction-intensive service on Ethereum as of Q2 2022 — to total transactions.
We also consider the timeout value for inference requests, $q.timeout$, with a default value of 20 blocks, which is approximately 4 minutes on the Ethereum.
Another variable of interest is the difficulty level $d_I$ for being elected as a member of the inference committee. Lowering $d_I$ reduces the probability of an individual being elected. We experimented with a fixed base difficulty of $2^{255}$, signifying that each member has a 50\% ($=2^{255}/2^{256}$) chance of being elected. For comparison, the $2^{253}$ used in the experiment has a 12.5\% ($=2^{253}/2^{256}$) chance of being elected.
We set the total number of BRAIN nodes to 21, inspired by the EOS blockchain's BFT-DPoS consensus mechanism's setting~\cite{eos}. This choice balances decentralization, governance, and scalability while maintaining network security~\cite{eosinterview}. Thus, the minimum number of nodes required to pass phase \textcircled{\raisebox{-0.9pt}{2}}-a, $Q_C$, has a maximum limit of 21. The default value of $Q_C$ is set to 11, which is approximately half and over 50\% of 21. Refer to section~\ref{subsubsection:tps-Qc} for a higher quorum value of 15, similar to EOS.

Other settings were controlled to ensure they did not affect the performance of the experiments.
The blockchain specifications used in the BRAIN platform were based on Ethereum, with a block interval of 12.06 seconds and an average of 155 transactions per block as of January 13, 2023.
To measure the performance of the platform, we quantified the execution time of non-inference transactions as 1 ms, based on the EVM transaction performance times reported in \cite{kim2021ethanos}. 
An epoch $E_I$ that changes the message used as input for the election of the inference committee was fixed at 8 blocks.
The duration of each phase \textcircled{\raisebox{-0.9pt}{2}} period is considered infinite.
Under the widely-accepted assumption in blockchain systems that more than half of the participating nodes are honest~\cite{nakamoto2008bitcoin}, the BRAIN's inference and learning system, which utilizes majority and median values, always operates correctly.
However, while the integrity of the results cannot be altered, malicious attacks that intentionally withhold committed outputs may affect the system's bandwidth. This can be measured by an increase in the quorum ratio proportional to the number of attackers, which in turn results in decreased bandwidth and increased timeouts. These metrics are discussed in the following section~\ref{subsubsection:tps-Qc}.
In this experiment, we consider $Q_R$ to be always equal to $Q_C$. This is because we can account for the existence of malicious nodes by increasing $Q_C$, and since no malicious nodes are present in our experiment, the assumption that $Q_R = Q_C$ represents the worst-case scenario, the minimum throughput. This robustly explains our results and provides a conservative estimate of the system's performance.

In our experiments, we used the float16 version of GPT-J-6B \cite{gpt-j}, a 6 billion parameter open-source version of GPT-3 \cite{brown2020language}. No optimization techniques were applied during inference.
All test datasets for inference were sourced from the SAMSum~\cite{gliwa2019samsum}. A total of 819 datasets were used for evaluation.
To determine the priority of each request when pushing to the priority queue, we sampled values along a Pareto distribution~\cite{pareto1897cours, arnold2014pareto} to simulate real-world $feePrice$ allocation. Priorities have a discrete range of 0 to 1000, with larger values indicating higher priority.

\bigskip

\subsubsection{\textbf{Inference Response Latency}}

To measure the performance of large neural networks separate from BRAIN, we conducted an experiment measuring the time required for inference on a single system consisting of one RTX 3090 graphics card, running Ubuntu 22.04.1.
This experiment was conducted 10 times using a fixed seed to obtain the average time spent.
The minimum inference time recorded was 0.0975 seconds, while the maximum time reached was 50.6394 seconds, resulting in an average time of 18.5421 seconds.

With the inference time for each request, we then implemented a simulator to measure the delay of BRAIN on a blockchain.
We set the $freq$ of the inference requests compared with general transactions as a default value of 0.0577, and other hyperparameters also used their default values.
The experiment was simulated 100 times to obtain the average value.
As a result, BRAIN required a minimum of 2 blocks, a maximum of 23 blocks, and an average of 7.1653 blocks ($\sigma=3.5919$) to receive a response after the request.
It is obvious that a minimum of 2 blocks is necessary because at least two transactions, \textsc{Commit} and \textsc{Reveal}, are mandated by the commit-and-reveal scheme. They must be located in different blocks since each BRAIN node can send a \textsc{Reveal} transaction only after recognizing the end of phase \textcircled{\raisebox{-0.9pt}{2}}-a.

Since Ethereum's block interval is 12.06 seconds, this corresponds to a minimum of 24.12 seconds and a maximum of 277.38 seconds, with an average of 86.4135 seconds.
By comparing this to the time required for inference on a single computer, we find that the time cost of reliably performing AI inference using BRAIN is approximately $67.8714=(86.4135-18.5421)$ seconds.
This value is heavily influenced by the quorum $Q_C$, and inference performance with different values of $Q_C$ can be observed in the following section~\ref{subsubsection:tps-Qc}.

\bigskip

\subsubsection{\textbf{Tasks Per Second}}

\begin{figure}
\centering
\begin{tikzpicture}
\centering
\begin{axis}[
    width=0.9\columnwidth,
    height=0.4\columnwidth,
    ybar,
    ymin=0,
    ymax=40,
    axis y line*=right,
    ylabel={\# of Timed out},
    ylabel style={red, font=\footnotesize},
    ytick=\empty,
    extra y ticks={0, 10, 20, 30, 40},
    yticklabel style={red, font=\footnotesize},
    symbolic x coords={
        {0.00005},
        {0.0001},
        {0.001},
        {0.005},
        {0.01},
        {0.05},
        {0.1}
    },
    xtick distance=1,
    xtick={
        {0.00005},
        {0.0001},
        {0.001},
        {0.005},
        {0.01},
        {0.05},
        {0.1}
    },
    xticklabels={
        {0.00005},
        {},
        {0.001},
        {0.005},
        {0.01},
        {0.05},
        {0.1}
    },
    xticklabel style = {
        font=\footnotesize,
    },
    extra x ticks={{0.0001}},
    extra x tick style={
        tick label style={
            xshift=0cm,
            yshift=.36cm
        },
    },
    extra x tick labels={\color{black}{/\! \!/}},
    extra x tick style=grid=major,
]
    \addplot+[
        red,fill=red!30!white,
    ][
        error bars/.cd,
        y dir=both,
        y explicit,
    ] coordinates {
        ({0.00005}, 0.0) +- (0, 0.0) 
        ({0.001}, 0.0) +- (0, 0.0) 
        ({0.005}, 0.0) +- (0, 0.0) 
        ({0.01}, 0.0) +- (0, 0.0) 
        ({0.05}, 0.6600) +- (0, 1.4712) 
        ({0.1}, 20.8200) +- (0, 9.1185) 
    };
\end{axis}

\begin{axis}[
    width=0.9\columnwidth,
    height=0.4\columnwidth,
    ymin=0,
    ymax=1050,
    axis x line=none,
    axis y line*=left,
    ylabel={Tasks Per Second},
    ylabel style={blue, font=\footnotesize},
    ytick=\empty,
    extra y ticks={0, 200, 400, 600, 800, 1000},
    yticklabel style={blue, font=\footnotesize},
    symbolic x coords={
        {0.00005},
        {0.0001},
        {0.001},
        {0.005},
        {0.01},
        {0.05},
        {0.1}
    },
    xtick distance=1,
    xtick={
        {0.00005},
        {0.001},
        {0.001},
        {0.005},
        {0.01},
        {0.05},
        {0.1}
    },
    xticklabels={
        {0.00005},
        {},
        {0.001},
        {0.005},
        {0.01},
        {0.05},
        {0.1}
    },
]
    \addplot[
        mark=*,mark options={fill=blue},
    ][
        error bars/.cd,
        y dir=both,
        y explicit,
    ] coordinates {
        ({0.00005}, 998.901) +- (0, 0.0) 
        ({0.001}, 978.4262) +- (0, 0.5304) 
        ({0.005}, 901.4241) +- (0, 0.1232) 
        ({0.01}, 821.2287) +- (0, 0.0822)
        ({0.05}, 490.7004) +- (0, 0.0241)
        ({0.1}, 339.4203) +- (0, 0.1367)
    };
    \label{graph:BRAIN}

    \addplot[
        mark=*,mark options={fill=blue},
        draw=none,
        nodes near coords,
        node near coords style={
            text=blue,
            font=\footnotesize, 
            /utils/exec={
                \setbox0\hbox{\pgfmathprintnumber\pgfplotspointmeta}
                \pgfmathfloattomacro{\pgfplotspointmeta}{\F}{\M}{\E}
                \pgfmathsetmacro{\myanchor}{
                    ifthenelse(\M*pow(10,\E-2)*5-\the\wd0>0,"north","south")
                }
            },
            anchor=north,
            rotate=20,
            xshift=-0.4cm,
            yshift=-0.03cm,
        },
    ][
        error bars/.cd,
        y dir=both,
        y explicit,
    ] coordinates {
        ({0.001}, 978.4712) +- (0, 0.0256) 
        ({0.005}, 901.3311) +- (0, 0.1193) 
        ({0.01}, 821.3025) +- (0, 0.2608)
    };

    \addplot[
        mark=*,mark options={fill=blue},
        draw=none,
        nodes near coords,
        node near coords style={
            text=blue,
            font=\footnotesize, 
            /utils/exec={
                \setbox0\hbox{\pgfmathprintnumber\pgfplotspointmeta}
                \pgfmathfloattomacro{\pgfplotspointmeta}{\F}{\M}{\E}
                \pgfmathsetmacro{\myanchor}{
                    ifthenelse(\M*pow(10,\E-2)*5-\the\wd0>0,"north","south")
                }
            },
            anchor=north,
            rotate=20,
            xshift=-0.55cm,
            yshift=0.2cm,
        },
    ][
        error bars/.cd,
        y dir=both,
        y explicit,
    ] coordinates {
        ({0.05}, 490.9023) +- (0, 0.4506)
    };

    \addplot[
        mark=*,mark options={fill=blue},
        draw=none,
        nodes near coords,
        node near coords style={
            text=blue,
            font=\footnotesize, 
            /utils/exec={
                \setbox0\hbox{\pgfmathprintnumber\pgfplotspointmeta}
                \pgfmathfloattomacro{\pgfplotspointmeta}{\F}{\M}{\E}
                \pgfmathsetmacro{\myanchor}{
                    ifthenelse(\M*pow(10,\E-2)*5-\the\wd0>0,"north","south")
                }
            },
            anchor=north,
            rotate=20,
            xshift=-0.67cm,
            yshift=0.24cm,
        },
    ][
        error bars/.cd,
        y dir=both,
        y explicit,
    ] coordinates {
        ({0.1}, 343.2634) +- (0, 3.3323)
    };
\end{axis}

\begin{axis}[
    width=0.9\columnwidth,
    height=0.4\columnwidth,
    ymin=0,
    ymax=1050,
    axis x line=none,
    axis y line=none,
    symbolic x coords={
        {0.00005},
        {0.0001},
        {0.001},
        {0.005},
        {0.01},
        {0.05},
        {0.1}
    },
    xtick distance=1,
    xticklabels={
        {0.00005},
        {},
        {0.001},
        {0.005},
        {0.01},
        {0.05},
        {0.1}
    },
    xtick={
        {0.00005},
        {0.0001},
        {0.001},
        {0.005},
        {0.01},
        {0.05},
        {0.1}
    },
]
    \addplot[mark=triangle*,mark options={fill=green}][
        error bars/.cd,
        y dir=both,
        y explicit,
    ] coordinates {
        ({0.00005}, 518.9390) +- (0, 0.0)
        ({0.001}, 51.5873) +- (0, 0.0594) 
        ({0.005}, 10.8143) +- (0, 0.0143) 
        ({0.01}, 5.4704) +- (0, 0.0096)
        ({0.05}, 1.1518) +- (0, 0.0000)
        ({0.1}, 0.6092) +- (0, 0.0042)
    };
    \label{graph:Naive}
\end{axis}

\begin{axis}[
    legend pos=north west,
    legend style={
        legend columns=-1,
        font=\footnotesize,
        at={(0.0,1.3)},
        anchor=west
    },
    width=0.9\columnwidth,
    height=0.4\columnwidth,
    ymin=0,
    ymax=1050,
    axis x line=none,
    axis y line=none,
    symbolic x coords={
        {0.00005},
        {0.0001},
        {0.001},
        {0.005},
        {0.01},
        {0.05},
        {0.1}
    },
    xtick distance=1,
    xticklabels={
        {0.00005},
        {},
        {0.001},
        {0.005},
        {0.01},
        {0.05},
        {0.1}
    },
    xtick={
        {0.00005},
        {0.0001},
        {0.001},
        {0.005},
        {0.01},
        {0.05},
        {0.1}
    },
]
    \addlegendimage{/pgfplots/refstyle=graph:BRAIN}\addlegendentry{BRAIN}
    \addlegendimage{/pgfplots/refstyle=graph:Naive}\addlegendentry{Naïve}
    \addplot[mark=diamond*,mark options={fill=orange}] coordinates {
        ({0.00005}, 1000.0)
        ({0.001}, 1000.0)
        ({0.005}, 1000.0)
        ({0.01}, 1000.0)
        ({0.05}, 1000.0)
        ({0.1}, 1000.0)
    };
    \addlegendentry{No Inference}
\end{axis}

\end{tikzpicture}
\caption{
    Graph illustrating the relationship between the frequency of inference requests with the number of tasks per second (line) and the number of timed-out requests (bar).
}
\label{fig:simul-graph}
\end{figure}
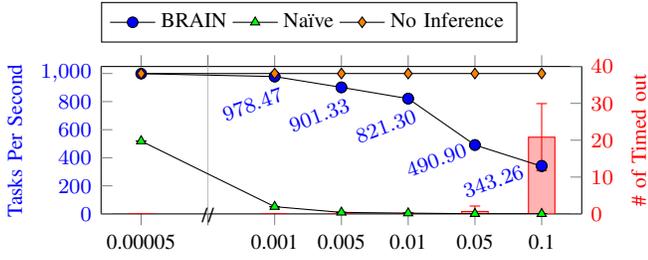

Since the inference operations in BRAIN are off-chain, there is no computational on-chain delay for inference transactions. Consequently, the Transactions-Per-Second (TPS) is measured at the same value ($1/0.001=1000$) as usual, given that we assumed the transaction execution time to be 1 ms.
To accurately evaluate the performance of BRAIN's decentralized inference system, we have defined the \textit{tasks-per-second} metric. This metric counts the number of tasks completed per second, grouping all transactions related to a single request as one task. Each task may consist of multiple transactions and takes the total execution time of those transactions to complete. Transactions unrelated to inference are considered separate tasks individually.

Fig.~\ref{fig:simul-graph} presents the average and standard deviation of 100 simulations for different $freq$ of inference requests.
The performance of BRAIN's tasks-per-second is illustrated in the line graph marked by a blue circle. As $freq$ increases, the tasks-per-second decreases since processing an inference request necessitates the aggregation of multiple transactions, including \textsc{Request} and all \textsc{Commit} and \textsc{Reveal} transactions, into a single task.
However, due to the BRAIN's pipelining ability, the decrease is not linear and is mitigated.

For comparison, we include results from a naïve implementation, which simply waits for a response from a large neural network on-chain without using a two-phase transaction structure.
As indicated in the line graph marked by a green triangle in Fig.~\ref{fig:simul-graph}, the naïve implementation exhibits significantly lower tasks-per-second, emphasizing the effectiveness of BRAIN.
When the activity level is set to $freq=0.0577$, BRAIN can process an average of 458.0193 ($\sigma=0.8641$) tasks-per-second. In contrast, the naïve one only achieves an average performance of 1.0079. This demonstrates that BRAIN can process tasks 454.4293 times faster than the naïve approach.

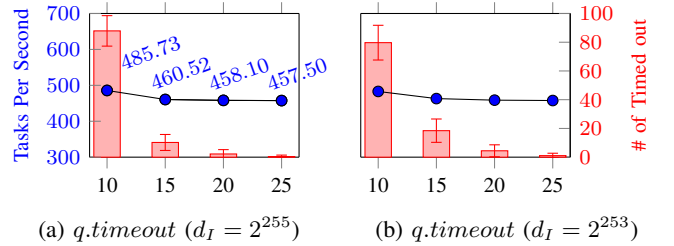
\begin{figure}
\centering
\begin{subfigure}{0.493\linewidth}
    \centering
    \begin{tikzpicture}

\begin{axis}[
    width=1.0\textwidth,
    height=0.8\textwidth,
    ybar,
    ymin=0,
    ymax=100,
    axis y line*=right,
    ytick=\empty,
    yticklabel=\empty,
    symbolic x coords={
        {10},
        {15},
        {20},
        {25}
    },
    xtick distance=1,
    xtick={
        {10},
        {15},
        {20},
        {25}
    },
    xticklabel style = {
        font=\footnotesize,
    },
]
    \addplot+[red,fill=red!30!white][
        error bars/.cd,
        y dir=both,
        y explicit,
    ] coordinates {
        ({10}, 87.8900) +- (0, 10.6065) 
        ({15}, 10.2300) +- (0, 5.5621) 
        ({20}, 2.2200) +- (0, 3.0285)
        ({25}, 0.4400) +- (0, 0.9932)
    };
\end{axis}

\begin{axis}[
    width=1.0\textwidth,
    height=0.8\textwidth,
    ymin=300,
    ymax=700,
    axis x line=none,
    axis y line*=left,
    ylabel={Tasks Per Second},
    ylabel style={blue, font=\footnotesize},
    ytick=\empty,
    extra y ticks={300, 400, 500, 600, 700},
    yticklabel style={blue, font=\footnotesize},
    symbolic x coords={
        {10},
        {15},
        {20},
        {25}
    },
    xtick distance=1,
    xtick={
        {10},
        {15},
        {20},
        {25}
    }
]
    \addplot[
        mark=*,mark options={fill=blue}
    ][
        error bars/.cd,
        y dir=both,
        y explicit,
    ] coordinates {
        ({10}, 485.7275) +- (0, 3.6290) 
        ({15}, 460.5220) +- (0, 1.7092) 
        ({20}, 458.0955) +- (0, 0.9233)
        ({25}, 457.5026) +- (0, 0.4811)
    };
    \addplot[
        nodes near coords,
        node near coords style={
            text=blue,
            font=\footnotesize, 
            xshift=0.65cm,
            yshift=0.3cm,
            rotate=20,
        },
        mark=*,mark options={fill=blue}
    ][
        error bars/.cd,
        y dir=both,
        y explicit,
    ] coordinates {
        ({10}, 485.7275) +- (0, 3.6290) 
    };
    \addplot[
        nodes near coords,
        node near coords style={
            text=blue,
            font=\footnotesize, 
            xshift=0.28cm,
            yshift=0.15cm,
            rotate=20,
        },
        mark=*,mark options={fill=blue}
    ][
        error bars/.cd,
        y dir=both,
        y explicit,
    ] coordinates {
        ({15}, 460.5220) +- (0, 1.7092) 
        ({20}, 458.0955) +- (0, 0.9233)
        ({25}, 457.5026) +- (0, 0.4811)
    };
\end{axis}
\end{tikzpicture}
    \caption{$q.timeout$ ($d_I=2^{255}$)}
    \label{fig:timeout-d-graph-1}
\end{subfigure}
\hfill
\begin{subfigure}{0.493\linewidth}
    \centering
    \begin{tikzpicture}

\begin{axis}[
    width=1.0\textwidth,
    height=0.8\textwidth,
    ybar,
    ymin=0,
    ymax=100,
    axis y line*=right,
    ylabel={\# of Timed out},
    ylabel style={red, font=\footnotesize},
    ytick=\empty,
    extra y ticks={0, 20, 40, 60, 80, 100},
    yticklabel style={red, font=\footnotesize},
    symbolic x coords={
        {10},
        {15},
        {20},
        {25}
    },
    xtick distance=1,
    xtick={
        {10},
        {15},
        {20},
        {25}
    },
    xticklabel style = {
        font=\footnotesize,
    },
]
    \addplot+[red,fill=red!30!white][
        error bars/.cd,
        y dir=both,
        y explicit,
    ] coordinates {
        ({10}, 79.7000) +- (0, 12.0876) 
        ({15}, 18.4400) +- (0, 8.1244) 
        ({20}, 4.4400) +- (0, 4.1577)
        ({25}, 1.0700) +- (0, 1.6689)
    };
\end{axis}

\begin{axis}[
    width=1.0\textwidth,
    height=0.8\textwidth,
    ymin=300,
    ymax=700,
    axis x line=none,
    yticklabel=\empty,
    ytick=\empty,
    symbolic x coords={
        {10},
        {15},
        {20},
        {25}
    },
    xtick distance=1,
    xtick={
        {10},
        {15},
        {20},
        {25}
    },
]
    \addplot[mark=*,mark options={fill=blue}][
        error bars/.cd,
        y dir=both,
        y explicit,
    ] coordinates {
        ({10}, 482.9529) +- (0, 4.0951) 
        ({15}, 463.0653) +- (0, 2.5606) 
        ({20}, 458.6965) +- (0, 1.3234)
        ({25}, 457.6686) +- (0, 0.6817)
    };
\end{axis}

\end{tikzpicture}
    \caption{$q.timeout$ ($d_I=2^{253}$)}
    \label{fig:timeout-d-graph-2}
\end{subfigure}
\caption{Tasks-per-second and the number of timeouts on various inference request's $q.timeout$. (a) shows the results at 50\% election probability and (b) at 12.5\%, based on $d_I$.}
\label{fig:timeout-d-graph}
\end{figure}

Furthermore, as the frequency of inference requests increases, a corresponding increase in the number of timed-out requests is observed, particularly beyond a threshold value of $freq=0.1$.
However, this issue can be mitigated by adjusting hyperparameters such as increasing $d_I$, as illustrated in Fig.~\ref{fig:qc-d-graph}, or increasing $q.timeout$ for each request, as depicted in Fig.~\ref{fig:timeout-d-graph}.

\bigskip

\subsubsection{\textbf{Tasks Per Second on various $Q_C$}} \label{subsubsection:tps-Qc}

\begin{figure}
\centering
\begin{subfigure}{0.493\linewidth}
    \centering
    \begin{tikzpicture}

\begin{axis}[
    width=1.0\textwidth,
    height=0.8\textwidth,
    ybar,
    ymin=0,
    ymax=100,
    axis y line*=right,
    yticklabel=\empty,
    ytick=\empty,
    symbolic x coords={
        {5},
        {10},
        {15},
        {21}
    },
    xtick distance=1,
    xtick={
        {5},
        {10},
        {15},
        {21}
    },
    xticklabel style = {
        font=\footnotesize,
    },
]
    \addplot+[
        red,fill=red!30!white
    ][
        error bars/.cd,
        y dir=both,
        y explicit,
    ] coordinates {
        ({5}, 1.5100) +- (0, 2.2338) 
        ({10}, 2.3200) +- (0, 3.3343) 
        ({15}, 2.1700) +- (0, 2.5419)
        ({21}, 4.3100) +- (0, 3.4805)
    };
\end{axis}

\begin{axis}[
    width=1.0\textwidth,
    height=0.8\textwidth,
    axis x line*=top,
    axis y line=none,
    yticklabel=\empty,
    ytick=\empty,
    symbolic x coords={
        {5},
        {10},
        {15},
        {21}
    },
    xtick distance=1,
    xtick={
        {5},
        {10},
        {15},
        {21}
    },
    xticklabels={
        {24\%},
        {48\%},
        {71\%},
        {100\%}
    },
    xticklabel style = {
        font=\footnotesize,
        yshift=0.13cm,
    },
]
    \addplot[
        draw=none,
    ] coordinates {
        ({5}, 1.5100)
        ({10}, 2.3200)
        ({15}, 2.1700)
        ({21}, 4.3100)
    };
\end{axis}

\begin{axis}[
    width=1.0\textwidth,
    height=0.8\textwidth,
    ymin=300,
    ymax=700,
    axis x line=none,
    axis y line*=left,
    ylabel={Tasks Per Second},
    ylabel style={blue, font=\footnotesize},
    yticklabel style={blue, font=\footnotesize},
    ytick=\empty,
    extra y ticks={300, 400, 500, 600, 700},
    symbolic x coords={
        {5},
        {10},
        {15},
        {21}
    },
    xtick distance=1,
    xtick={
        {5},
        {10},
        {15},
        {21}
    },
]
    \addplot[
        mark=*,mark options={fill=blue}
    ][
        error bars/.cd,
        y dir=both,
        y explicit,
    ] coordinates {
        ({5}, 649.9809) +- (0, 0.8270) 
        ({10}, 481.8634) +- (0, 1.0241) 
        ({15}, 382.6702) +- (0, 0.7372)
        ({21}, 307.4273) +- (0, 0.8958)
    };
    \addplot[
        draw=none,
        mark=*,mark options={fill=blue},
        nodes near coords,
        node near coords style={
            text=blue,
            font=\footnotesize, 
            xshift=0.25cm,
            yshift=0.15cm,
            rotate=20,
        },
    ][
        error bars/.cd,
        y dir=both,
        y explicit,
    ] coordinates {
        ({10}, 481.8634) +- (0, 1.0241) 
        ({15}, 382.6702) +- (0, 0.7372)
        ({21}, 307.4273) +- (0, 0.8958)
    };
    \addplot[
        draw=none,
        mark=*,mark options={fill=blue},
        nodes near coords,
        node near coords style={
            text=blue,
            font=\footnotesize, 
            xshift=0.7cm,
            yshift=-0.25cm,
            rotate=20,
        },
    ][
        error bars/.cd,
        y dir=both,
        y explicit,
    ] coordinates {
        ({5}, 649.9809) +- (0, 0.8270) 
    };
\end{axis}
\end{tikzpicture}
    \caption{$Q_C$ ($d_I=2^{255}$)}
    \label{fig:qc-d-graph-1}
\end{subfigure}
\hfill
\begin{subfigure}{0.493\linewidth}
    \centering
    \begin{tikzpicture}

\begin{axis}[
    width=1.0\textwidth,
    height=0.8\textwidth,
    ybar,
    ymin=0,
    ymax=100,
    axis y line*=right,
    ylabel={\# of Timed out},
    ylabel style={red, font=\footnotesize},
    ytick=\empty,
    extra y ticks={0, 20, 40, 60, 80, 100},
    yticklabel style={red, font=\footnotesize},
    symbolic x coords={
        {5},
        {10},
        {15},
        {21}
    },
    xtick distance=1,
    xtick={
        {5},
        {10},
        {15},
        {21}
    },
    xticklabel style = {
        font=\footnotesize,
    },
]
    \addplot[
        red,fill=red!30!white,
    ][
        error bars/.cd,
        y dir=both,
        y explicit,
    ] coordinates {
        ({5}, 1.7600) +- (0, 2.6043) 
        ({10}, 2.9700) +- (0, 2.6850) 
        ({15}, 18.5300) +- (0, 5.8180)
        ({21}, 274.3700) +- (0, 9.6121)
    };
    
    \node[
        rotate=90,
        font={
            \footnotesize
            \color{red}
        }
    ] at ({21}, 65) {
        274.3700
    };
\end{axis}

\begin{axis}[
    width=1.0\textwidth,
    height=0.8\textwidth,
    axis x line*=top,
    axis y line=none,
    yticklabel=\empty,
    ytick=\empty,
    symbolic x coords={
        {5},
        {10},
        {15},
        {21}
    },
    xtick distance=1,
    xtick={
        {5},
        {10},
        {15},
        {21}
    },
    xticklabels={
        {24\%},
        {48\%},
        {71\%},
        {100\%}
    },
    xticklabel style = {
        font=\footnotesize,
        yshift=0.13cm,
    },
]
    \addplot[
        draw=none,
    ] coordinates {
        ({5}, 1.5100)
        ({10}, 2.3200)
        ({15}, 2.1700)
        ({21}, 4.3100)
    };
\end{axis}

\begin{axis}[
    width=1.0\textwidth,
    height=0.8\textwidth,
    ymin=300,
    ymax=700,
    axis x line=none,
    yticklabel=\empty,
    ytick=\empty,
    symbolic x coords={
        {5},
        {10},
        {15},
        {21}
    },
    xtick distance=1,
    xtick={
        {5},
        {10},
        {15},
        {21}
    },
]
    \addplot[mark=*,mark options={fill=blue}][
        error bars/.cd,
        y dir=both,
        y explicit,
    ] coordinates {
        ({5}, 650.1909) +- (0, 0.7268) 
        ({10}, 482.0358) +- (0, 0.8233) 
        ({15}, 387.0812) +- (0, 1.8722)
        ({21}, 401.0228) +- (0, 4.1204)
    };
\end{axis}

\end{tikzpicture}
    \caption{$Q_C$ ($d_I=2^{253}$)}
    \label{fig:qc-d-graph-2}
\end{subfigure}
\caption{
    Tasks-per-second and the number of timeouts on various $Q_C$.
    The above x-axis represents the percentage of $Q_C$ to the total number of nodes.
    (a) shows the results at 50\% VRF election probability and (b) at 12.5\%, depending on $d_I$.
}
\label{fig:qc-d-graph}
\end{figure}
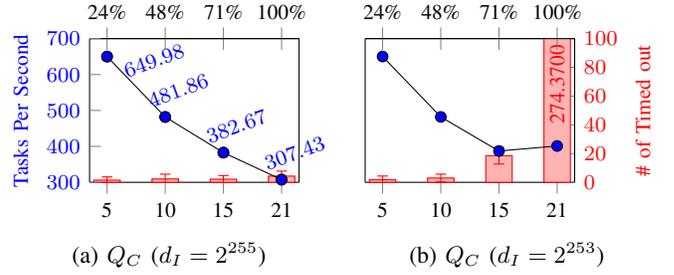

Fig.~\ref{fig:qc-d-graph} illustrates the experimental results of the tasks-per-second and the number of timeouts in relation to changes in quorum $Q_C$, which is the number of quorum in phase \textcircled{\raisebox{-0.9pt}{2}}-a.
As the number of required nodes in one inference increases, the overall transaction requirement generated by the inference committee likewise increases, leading to a decrease in the tasks-per-second.
Conversely, a smaller quorum leads to higher performance.
However, it is important to note that balancing the trade-off between security and performance is key when setting the quorum size, as reducing it can increase performance but also threaten the reliability of inference results.

As shown in Fig.~\ref{fig:qc-d-graph-1}, increasing the quorum slightly increases the number of timeouts, but BRAIN can still handle a large number of quorums in an environment with default values of $E_I=8$ and $q.timeout=20$.
However, as shown in Fig.~\ref{fig:qc-d-graph-2}, in a difficult $2^{253}$, the number of timeouts increases significantly at a quorum level of 15, reaching an average of 274.37 timeouts at 21, as the platform is broken under this hyperparameter setting.
Adjusting the difficulty $d_I$ instead of $Q_C$ can decrease the number of timeouts, but this in turn decreases the computational resource efficiency and increases the cost per inference.
Therefore, when setting hyperparameters such as $T_C$, $E_I$, $d_I$, $q.timeout$, and $Q_C$, it is important to determine an appropriate range that the BRAIN can handle based on the desired level of performance, security, and timeout tolerance under the expected service activation degree $freq$.

\bigskip

\subsubsection{\textbf{Number of Timeouts on various $q.timeout$}}

The timeout value included in the request corresponds to the validity period, measured in blocks.
Fig.~\ref{fig:timeout-d-graph} demonstrates that shortening the $q.timeout$ increases the likelihood of more requests exceeding their expiration date, particularly when the value falls below a certain threshold.
In the default value environment, timeouts become severe when $q.timeout \le 10$.

Meanwhile, Fig.~\ref{fig:timeout-d-graph-1} and Fig.~\ref{fig:timeout-d-graph-2} compare experimental results for difficulty levels of $2^{255}$ and $2^{253}$, respectively.
Although decreasing the probability of being selected as a VRF with a fixed quorum might increase the likelihood of timeouts due to longer quorum fulfillment times, the experiments show only a small difference.
This indicates that under appropriate settings for $E$, $Q_C$, and $freq$, BRAIN can handle changes in $d_I$ within an acceptable range. Conversely, when the $Q_C$ and $d_I$ settings are inappropriate, the number of timeouts increases dramatically, as shown in Fig.~\ref{fig:qc-d-graph-2}.

\subsection{Contract Overhead Analysis} \label{subsection:contract-overhead}

We analyze the cost overhead incurred by various techniques used in BRAIN to ensure authenticity, security, and low latency. Specifically, we examine the \textbf{verifiable random function}, the \textbf{commit-and-reveal} scheme, and the \textbf{queue}.
We evaluate the associated gas consumption and corresponding costs on the Ethereum~\cite{wood2014ethereum} and Polygon~\cite{polygon} in Table~\ref{tab:vrf-gas}.
This provides a quantitative analysis of the overhead incurred by these techniques.
We have used the gas price of 14 gwei ($14 * 10^{-9}$ ETH) and 51.6 gwei ($51.6 * 10^{-9}$ MATIC) on Ethereum and Polygon respectively, as of January 13, 2023.

\begin{table}[!t]
\caption{Gas Consumption on Ethereum and Polygon}
\centering
\resizebox{\columnwidth}{!}{%
\begin{tabular}{l|r|r|r|r|r}
    \hline
        \multicolumn{1}{c|}{\multirow{2}{*}{Methods}}
        & \multicolumn{1}{c|}{\multirow{2}{*}{min}}
        & \multicolumn{1}{c|}{\multirow{2}{*}{max}}
        & \multicolumn{1}{c|}{\multirow{2}{*}{avg}}
        & \multicolumn{2}{c}{USD (avg)} \\\cline{5-6}
        \multicolumn{1}{c|}{} & & & & Ethereum & Polygon\\

    \hline \hline
        \rowcolor{Gray} \multicolumn{6}{c}{Verifiable Random Function} \\
    \hline
        \texttt{verify} & 1543493 & 1862450 & 1643712 & \$ 32.504 & \$ 0.077 \\
    \hline
        \texttt{verify}$_{fast}$ & 106360 & 352838 & 150715 & \$ 2.980 & \$ 0.007 \\
    \hline \hline
        \rowcolor{Gray} \multicolumn{6}{c}{Commit-and-Reveal: without \& with the Hash Function $H$} \\
    \hline
        \textsc{commit} & 44825 & 62072 & 45732
        & \$ 0.904 & \$ 0.002
        \\
    \hline
        \textsc{commit}$_{H}$ & 44861 & 44897 & 44895
        & \$ 0.888 & \$ 0.002
        \\
    \hline
        \textsc{reveal} & 27831 & 796620 & 87124
        & \$ 1.723 & \$ 0.004
        \\
    \hline
        \textsc{reveal}$_{H}$ & 47355 & 47391 & 47389
        & \$ 0.937 & \$ 0.002
        \\
    \hline \hline
        \rowcolor{Gray} \multicolumn{6}{c}{(Circular) Queue \& Priority Queue} \\
    \hline
        \texttt{push} & 51324 & 68424 & 51345 & \$ 1.015 & \$ 0.002 \\
    \hline
        \texttt{pop} & 29013 & 46113 & 29034 & \$ 0.574 & \$ 0.001 \\
    \hline
        \texttt{push}$_{prior}$ & 84353 & 137955 & 91699 & \$ 1.813 & \$ 0.004 \\
    \hline
        \texttt{pop}$_{prior}$ & 34909 & 116942 & 100861 & \$ 1.995 & \$ 0.005 \\
    \hline \hline
        \multicolumn{4}{>{\columncolor[gray]{0.8}}l|}{\textcircled{\raisebox{-0.9pt}{1}} \texttt{push}$_{prior} =$}
        & \$ 1.813 & \$ 0.004
        \\
    \hline
        \multicolumn{4}{>{\columncolor[gray]{0.8}}l|}{\textcircled{\raisebox{-0.9pt}{2}}-a. \texttt{verify}$_{fast}$ + \texttt{commit}$_{H}$ + \texttt{pop}$_{prior} =$}
        & \$ 5.863 & \$ 0.014
        \\
    \hline
        \multicolumn{4}{>{\columncolor[gray]{0.8}}l|}{\textcircled{\raisebox{-0.9pt}{2}}-b. \texttt{reveal}$_{H} =$}
        & \$ 0.937 & \$ 0.002
        \\
    \hline
\end{tabular}%
}
\label{tab:vrf-gas}
\end{table}


Regarding the VRF, we observed that using the \texttt{ecrecover}-based \texttt{verify}$_{fast}$ function significantly reduces the cost to an average of \$2.98 per verification, compared with \$32.5 for full verification on Ethereum.

Concerning the commit-and-reveal scheme, we measured the cost of \textsc{Commit} and \textsc{Reveal} transactions for 819 actual inference results on the SAMSum \cite{gliwa2019samsum} test dataset.
We trivially observed that the difference in gas consumption between using the hash function on inference $output$ during \textsc{Commit} or not was insignificant.
However, there was a noticeable difference in gas consumption between using and not using the hash function during \textsc{Reveal}, since using the hashed value in the transaction uniformizes the gas consumption.
The overhead cost associated with the commit-and-reveal scheme in Ethereum is considered practical, with around \$1 to \$2.

We measured the gas cost for $push$ and $pop$ operations to evaluate the efficiency of the circular queue and priority queue. The results showed that the priority queue uses slightly more gas than the regular queue. However, overall, the overhead cost on Ethereum is low, ranging from \$0.5 to \$2.0.

\section{Applications} \label{section:applications}

BRAIN offers potential applications across various industries by providing security, traceability, and decentralization.


\vspace{4pt}

\noindent \textbf{Unbiased Generative AI.}
Generative AI models have gained significant attention due to their ability to create realistic, high-quality content~\cite{brown2020language, Midjourney, agostinelli2023musiclm, rombach2022highresolution}.
Despite their potential for creative expression, they present challenges regarding potential biases from training data.
To address this issue, BRAIN offers a decentralized training approach, enabling aggregator-free federated learners to join the training and evaluate the model based on their private dataset, considered the most trustworthy.

\vspace{4pt}

\noindent \textbf{Recommendation System.}
AI-driven content recommendations are widely used, including related movie, video, and music suggestions~\cite{cheng2016wide, wu2022graph}.
However, users often can't determine whether these recommendations result from genuine AI inferences or manipulated advertisements.
With a recommendation system on BRAIN that leverages the traceability of blockchain, users can verify that the suggested content originates from actual inferences.

\vspace{4pt}

\noindent \textbf{Intelligent NFT.}
Non-fungible tokens (NFTs) \cite{eip721} have emerged as a popular means of representing unique digital assets on the blockchain, often used for avatars and profile pictures~\cite{casale2022impact, bayc}.
By integrating the BRAIN, NFT creators can embed AI models within their digital assets, allowing for intelligent capabilities like dynamic animations and chat.
The decentralized nature of BRAIN ensures that these AI-powered NFTs are not subject to centralized control.

\section{Conclusion} \label{section:conclusion}

BRAIN is a contract-driven decentralized platform designed to enable the training and inference of large-scale neural networks for a variety of AI-based decentralized services.
To ensure reliability and security, BRAIN employs a two-phase transaction structure and a VRF-based committee sortition mechanism.
Additionally, we propose an aggregator-free federated learning mechanism that eliminates the need for a centralized server, providing cost efficiency.

Our experiments have showcased the effectiveness of these features in terms of tasks-per-second performance, achieving a 454.4293 times improvement compared to a naïve implementation.
By fine-tuning the hyperparameters, we successfully strike a balance between performance, security, and timeout tolerance, while effectively controlling the number of timeouts and tasks-per-second.
Moreover, we have demonstrated that BRAIN incurs a low additional gas cost overhead.

Note that, at present, BRAIN does not include privacy-preserving features, and this is not the focus of the current paper. In future research, we plan to explore methods for enhancing privacy on the BRAIN platform.

\section*{Acknowledgment}

This work was supported partly by Kakao Brain corporations.

\bibliographystyle{plain}
\bibliography{reference}

\end{document}